\documentclass{ws-ijmpcs}

\begin{document}

\markboth{M. L. L. da Silva, D, Hadjimichef and C. A. Z. Vasconcellos}
{Glueball-glueball potential in a constituent gluon model}

%
\catchline{}{}{}{}{}
%

\title{GLUEBALL-GLUEBALL POTENTIAL IN A CONSTITUENT GLUON MODEL}

\author{M. L. L. DA SILVA}

\address{Instituto de F\'{\i}sica e Matem\'atica, Universidade Federal de Pelotas \\ Caixa Postal 354, CEP 96010-090, Pelotas, RS, Brazil.\\
mario.silva@ufpel.edu.br}

\author{D. HADJIMICHEF}

\address{Instituto de F\'{\i}sica, Universidade Federal do Rio Grande do Sul \\ Caixa Postal 15051, CEP 91501-970, Porto Alegre, RS, Brazil.\\
dimiter.hadjimichef@ufrgs.br}

\author{C. A. Z. VASCONCELLOS}

\address{Instituto de F\'{\i}sica, Universidade Federal do Rio Grande do Sul \\ Caixa Postal 15051, CEP 91501-970, Porto Alegre, RS, Brazil.\\
cesarzen10@hotmail.com}

\maketitle

\begin{history}
\received{Day Month Year}
\revised{Day Month Year}
\end{history}

\begin{abstract}
In this work we use a mapping technique to derive in the context of a constituent gluon model an effective
Hamiltonian that involves explicit gluon degrees of freedom. We study glueballs with two gluons using the
Fock-Tani formalism. In the present work we calculate the glueball-glueball potential, in the context of the
constituent gluon model, with gluon interchange. 

\end{abstract}

\ccode{PACS numbers: 12.39.Mk,12.39.Pn, 12.39.Jh}

\section{Introduction}

The gluon self-coupling in QCD implies the existence of bound states of pure gauge
fields known as glueballs. Numerous technical difficulties have so far been present
in our understanding of their properties in experiments, largely because glueball states
can mix strongly with nearby $q\bar{q}$ resonances. In the present we follow a
different approach by applying the Fock-Tani formalism in order to obtain an effective interaction
between glueballs \cite{jphysg} and then a glueball-glueball potential can be obtained.

\section{The Fock-Tani Formalism}	

\begin{figure}[ht]
\begin{center}
\centerline{\psfig{file=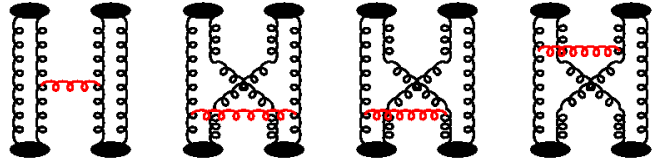,width=9cm}}
\end{center}
\caption{Diagrams representing the scattering amplitude $h_{fi}$ for
  glueball-glueball interaction with constituent gluon interchange.}
\label{hfi_fig}
\end{figure}

The starting point in the present calculation is  the definition, in second quantization, of the
glueball creation operator formed by two constituent gluons 
\begin{eqnarray}
    G_{\alpha}^{\dagger}=\frac{1}{\sqrt{2}}\Phi_{\alpha}^{\mu \nu}
    a_{\mu}^{\dagger}a_{\nu}^{\dagger}.
\label{G}
\end{eqnarray}
Gluon creation $a^{\dag}_{\nu}$ and annihilation $a_{\mu}$ operators obey the conventional
commutation relations
\begin{eqnarray}
    [a_{\mu},a_{\nu}]=0 \,\,\,\, ; \,\,\,\,[a_{\mu},a_{\nu}^{\dagger}]=\delta_{\mu\nu}.
\end{eqnarray}
In (\ref{G}) $\,\Phi_{\alpha}^{\mu\nu}$ is the bound-state wave-function for two-gluons. The
composite glueball operator satisfy non-canonical commutation relations
\begin{eqnarray}
[G_{\alpha},G_{\beta}]=0\,\,\,\,\,;\,\,\,\, [G_{\alpha},G_{\beta}^{\dagger}]
    =\delta_{\alpha\beta}+\Delta_{\alpha\beta}
\end{eqnarray}
where
\begin{eqnarray}
    \delta_{\alpha\beta}=\Phi_{\alpha}^{\star\rho \gamma}\Phi_{\beta}^{\gamma \rho}
\,\,\,\,\,\,;\,\,\,\,\,\,
    \Delta_{\alpha\beta}=2\Phi_{\alpha}^{\star\mu
      \gamma}\Phi_{\beta}^{\gamma \rho}a_{\rho}^{\dagger}a_{\mu}.
\end{eqnarray}
The  ``ideal particles'' which obey canonical relations, in our case  are the ideal glueballs
\begin{eqnarray}
    [g_{\alpha},g_{\beta}]=0\,\,\,\,\,;\,\,\,\, [g_{\alpha},g_{\beta}^{\dagger}]
    =\delta_{\alpha\beta}.
\end{eqnarray}
This way one can transform the composite glueball state $|\alpha\rangle$  into an ideal state
$|\alpha\,)$ by
\begin{eqnarray*}
 |\alpha\,)=   U^{-1}(-\frac{\pi}{2})\,G_{\alpha}^{\dagger}
\,|0\rangle=g_{\alpha}^{\dagger}\,| 0\rangle
\end{eqnarray*}
where $ U=\exp({tF})$ and $F$ is the generator of the glueball transformation given by
\begin{eqnarray}
    F=g_{\alpha}^{\dagger}\tilde{G}_{\alpha}-\tilde{G}_{\alpha}^{\dagger}g_{\alpha}
    \label{F}
\end{eqnarray}
with
\begin{eqnarray*}
    \tilde{G}_{\alpha}=G_{\alpha}-\frac{1}{2}\Delta_{\alpha\beta}G_{\beta}
    -\frac{1}{2}G_{\beta}^{\dagger}[\Delta_{\beta\gamma},G_{\alpha}]G_{\gamma}.
\end{eqnarray*}
In order to obtain the effective glueball-glueball potential one has to use (\ref{F}) in a set of
Heisenberg-like equations for the basic operators $g,\tilde{G},a$
\begin{eqnarray*}
    \frac{dg_{\alpha}(t)}{dt}=[g_{\alpha},F]=\tilde{G}_{\alpha}\,\,\,\,\,;\,\,\,\,\,
    \frac{d\tilde{G}_{\alpha}(t)}{dt}=[\tilde{G}_{\alpha}(t),F]=-g_{\alpha}\,.
\end{eqnarray*}
The simplicity of these equations are not present in the equations for $a$
\begin{eqnarray*}
    \frac{da_{\mu}(t)}{dt}=[a_{\mu},F]=&-&\sqrt{2}\Phi_{\beta}^{\mu\nu}a_{\nu}^{\dagger}g_{\beta}
    +\frac{\sqrt{2}}{2}\Phi_{\beta}^{\mu\nu}a_{\nu}^{\dagger}\Delta_{\beta\alpha}g_{\beta}\\
    &+&\Phi_{\alpha}^{\star\mu\gamma}\Phi_{\beta}^{\gamma\mu^{'}}
    (G_{\beta}^{\dagger}a_{\mu^{'}}g_{\beta}-g_{\beta}^{\dagger}a_{\mu^{'}}G_{\beta})\\
    &-&\sqrt{2}(\Phi_{\alpha}^{\mu\rho^{'}}\Phi_{\rho}^{\mu^{'}\gamma^{'}}
    \Phi_{\gamma}^{\star\gamma^{'}\rho^{'}}\\
&&+\Phi_{\alpha}^{\mu^{'}\rho^{'}}\Phi_{\rho}^{\mu\gamma^{'}}
    \Phi_{\gamma}^{\star\gamma^{'}\rho^{'}})G_{\gamma}^{\dagger}a_{\mu^{'}}^{\dagger}
    G_{\beta}g_{\beta}.
\end{eqnarray*}
The solution for these equation can be found order by order in the wave-functions. For zero
order one has $a_{\mu}^{(0)}=a_{\mu}$
\begin{eqnarray*}
    g_{\alpha}^{(0)}(t)&=&G_{\alpha}\sin{t}+g_{\alpha}\cos{t}\\
G_{\beta}^{(0)}(t) &=&G_{\beta}\cos{t}-g_{\beta}\sin{t}.
\end{eqnarray*}
In the first order $g_{\alpha}^{(1)}=0,\,\,\,G_{\beta}^{(1)}=0$ and
\begin{eqnarray*}
    a_{\mu}^{(1)}(t)=\sqrt{2}\Phi_{\beta}^{\mu\nu}a_{\nu}^{\dagger}[G_{\beta}^{(0)}-G_{\beta}].
\end{eqnarray*}
For more details about this calculation see Ref \cite{jphysg}. The glueball-glueball potential
can be obtained applying in a standard way the Fock-Tani transformed operators to the
microscopic Hamiltonian
\begin{eqnarray*}
    {\cal{H}}(\mu\nu;\sigma\rho)=T_{\rm aa}(\mu)a_{\mu}^{\dagger}a_{\mu}+\frac{1}{2}
    V_{\rm aa}(\mu\nu;\sigma\rho)a_{\mu}^{\dagger}a_{\nu}^{\dagger}a_{\rho}a_{\sigma}
\end{eqnarray*}
where in this microscopic Hamiltonian $T_{\rm aa}$ is the kinetic energy and $V_{\rm aa}$ is
the  potential in the constituent model. After transforming  $ {\cal{H}}(\mu\nu;\sigma\rho)$ one
obtains for the glueball-glueball potential $V_{gg}$
\begin{eqnarray}
    V_{gg}=\sum_{i=1}^{4}V_{i}(\alpha\gamma;\delta\beta)g_{\alpha}^{\dagger}
    g_{\gamma}^{\dagger}g_{\delta}g_{\beta}
\label{v_gg}
\end{eqnarray}
and
\begin{eqnarray}
    &V_{1}(\alpha\gamma;\delta\beta)=2V_{aa}(\mu\nu;\sigma\rho)\Phi_{\alpha}^{\star\mu\tau}
    \Phi_{\gamma}^{\star\nu\xi}\Phi_{\delta}^{\rho\xi}\Phi_{\beta}^{\sigma\tau}\nonumber\\
    &V_{2}(\alpha\gamma;\delta\beta)=2V_{aa}(\mu\nu;\sigma\rho)\Phi_{\alpha}^{\star\mu\tau}
    \Phi_{\gamma}^{\star\nu\xi}\Phi_{\delta}^{\rho\tau}\Phi_{\beta}^{\sigma\xi}\nonumber\\
    &V_{3}(\alpha\gamma;\delta\beta)=V_{aa}(\mu\nu;\sigma\rho)\Phi_{\alpha}^{\star\mu\nu}
    \Phi_{\gamma}^{\star\lambda\xi}\Phi_{\delta}^{\sigma\lambda}\Phi_{\beta}^{\rho\xi}\nonumber\\
    &V_{4}(\alpha\gamma;\delta\beta)=V_{aa}(\mu\nu;\sigma\rho)\Phi_{\alpha}^{\star\mu\xi}
    \Phi_{\gamma}^{\star\nu\lambda}\Phi_{\delta}^{\lambda\xi}\Phi_{\beta}^{\rho\sigma}\,.
\label{v1-v4}
\end{eqnarray}
The scattering $T$-matrix is related directly to Eq. (\ref{v_gg}) 
\begin{eqnarray}
T(\alpha\beta;\gamma\delta)=(\alpha\beta|V_{gg}|\gamma\delta)\,.
\label{t-matrix}
\end{eqnarray}
Due to translational invariance, the $T$-matrix element is written as a momentum conservation
delta-function, times a Born-order matrix element, $h_{fi}$:
\begin{eqnarray}
T(\alpha\beta;\gamma\delta)=\delta^{(3)}( \vec{P}_{f}-\vec{P}_{i})\,h_{fi}
\label{t-hfi}
\end{eqnarray}
where $\vec{P}_{f}$ and  $\vec{P}_{i}$ are the final and initial momenta of the two-glueball
system. This result can be used in order to evaluate  the glueball-glueball potential 
\begin{eqnarray}
V_{GG}(r)=\int\,d^{3}Q\, e^{i\vec{Q}\cdot \vec{r}}\,h_{fi}(\vec{Q}).
\label{vgg}
\end{eqnarray}
The scattering amplitude $h_{fi}$ can be visualized in Fig. \ref{hfi_fig}.

\begin{figure}[htb]
\begin{center}
{\epsfig{file=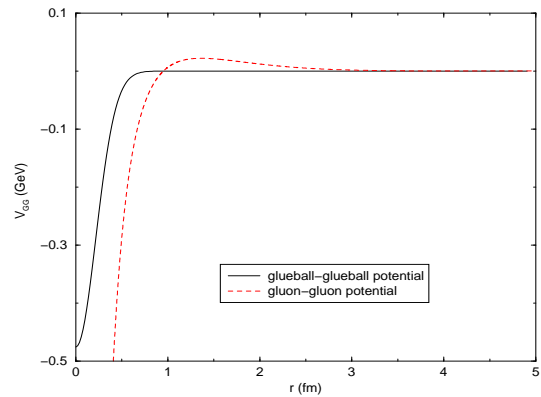, height=200.00pt, width=150.00pt,angle=-90}}
\end{center}
\caption{Glueball-glueball and gluon-gluon potential for $0^{++}$ with the following parameters
$r_0 = 0.58 \,fm$, $\beta=0.1$, $\lambda = 1.8$, $k=0.21$, gluon mass $m=0.6$ GeV. The  $s\bar{s}$ quark 
model parameters: $m_{q}=0.55$ GeV, $\alpha_{s}=0.6$.}
\label{graf_5}
\end{figure}

The potential $V_{\rm aa}$ is determined in the Cornwall and Soni constituent gluon
model \cite{cs1}
\begin{eqnarray}
V_{\rm aa}(r)= \frac{1}{3}\, f^{ace}f^{bde}   \left[\,V_{2g}^{OGEP}(r)
+V_{S}(r)
\right]
\label{mod24}
\end{eqnarray}
where
\begin{eqnarray}
V_{2g}^{OGEP}(r)=   -\lambda\,
 \left[ \,\omega_{1}\frac{e^{-m r}}{r}
+\omega_{2}\frac{\pi}{m^2}\,D(r)\,
\right]\,\,\,\,\,\,,\,\,\,\,\,\,
V_{S}(r)= 2m\,(1-e^{-\beta \,m\,r})
\label{mod24a}
\end{eqnarray}
and
\begin{eqnarray}
D(r)=\frac{k^{3}m^{3}}{\pi^{3/2}}\,e^{-k^{2}m^{2}r^{2}}
\,\,\,\,\,\,,\,\,\,\,\,\,
\lambda = \frac{3\,g^{2}}{4\pi} 
\,\,\,\,\,\,,\,\,\,\,\,\,
\omega_{1} = \frac{1}{4}+\frac{1}{3}\vec{S}^2 
\,\,\,\,\,\,,\,\,\,\,\,\,
\omega_{2} = 1-\frac{5}{6}\vec{S}^2  .
\label{mod24b}
\end{eqnarray}
The parameters $\lambda$, $m$, $k$ and $\beta$ assume known values\cite{cs2,cs3,cs4}
while  the wave function $\Phi^{\mu\nu}_{\alpha}$ is given in \cite{jphysg}.
The glueball-glueball scattering amplitude $h_{fi}$  is given by
\begin{eqnarray}
  h_{fi}(s,t) &=& \frac{3}{8}\,R_{0}(s)
\sum_{i=1}^{6}\,R_{i}(s,t)
\label{hfi-st}
\end{eqnarray}
where
\begin{eqnarray}
R_{0}&=& \frac{4}{(2\pi)^{3/2}b^3}
  \exp\left [-\frac{1}{2b^2} \left (\frac{s}{4} -M_{G}^2 \right)  \right]\,,\nonumber\\
R_{1}&=&\frac{\lambda \omega_{1}^{(2)}\, 4\,\sqrt{2\pi}}{3     }  \,
  \int_{0}^{\infty} dq\,\frac{q^2}{q^2 +m^{2}}
  \exp\left(-\frac{q^2}{2b^2}\right)
\left[
{\mathcal{J}}_0  \left(\frac{q\sqrt{t}}{2b^2}\right) +
{\mathcal{J}}_0  \left(\frac{q\sqrt{u}}{2b^2}\right)
\right]\,, \nonumber\\
R_{2}&=&
\frac{  \lambda \omega_{2}^{(2)}     2\sqrt{2} \pi b^3 k^3 m}{3(b^2
  +2k^2 m^{2})^{3/2}} 
\left[
\exp \left( -\frac{t k^2 m^2}{4(b^4 +2b^2 k^2 m^{2})}\right)
+\exp \left( -\frac{u k^2 m^2}{4(b^4 +2b^2 k^2 m^{2})}\right)
\right]\,, \nonumber\\
R_{3} &=&
\frac{32\sqrt{2\pi}}{3} \int_0^{\infty} dq\,
  \frac{q^2 \beta  m^{2}}{(q^{2} +\beta^{2}m^{2})^{2}} \exp
  \left(-\frac{q^2}{2b^2}\right)
\left[
 {\mathcal{J}}_0 \left(\frac{q\sqrt{t}}{2b^2}  \right)+
 {\mathcal{J}}_0 \left(\frac{q\sqrt{u}}{2b^2}  \right)
\right]\,, \nonumber\\
R_{4}&=&
-\frac{\lambda \omega_{1}^{(3)}}{3}\,\,
  \frac{16\,\sqrt{2\pi}\,b^2}{\sqrt{\frac{s}{4}-M_{G}^2}}
 \int_{0}^{\infty} dq\,\frac{q}{q^2 +m^{2}}
  \exp\left(-\frac{3q^2}{8b^2}\right)
  \sinh\left(\frac{q}{2b^2}\sqrt{\frac{s}{4}-M_{G}^2}\, \right) \,, \nonumber\\
R_{5}&=&-\frac{\lambda \omega_{2}^{(3)}\,}{3}\, \frac{16\pi b^3 k^3 m}{(2b^2
  +3k^2 m^{2})^{3/2}} \exp \left[ -\frac{k^2 m^2 \left( \frac{s}{4}
  -M_{G}^2\right)}{2(2b^4 +3b^2 k^2 m^{2})}\right]\,,\nonumber\\
R_{6}&=&-\frac{128 \sqrt{2\pi}
  b^2}{3\sqrt{\frac{s}{4} -M_{G}^2}} \int_0^{\infty} dq\,
  \frac{q \beta  m^{2}}{(q^{2} +\beta^{2}m^{2})^{2}}
  \exp \left(\frac{3q^2}{8b^2}\right)  \sinh \left(\frac{q}{2b^2}
  \sqrt{\frac{s}{4}-M_{G}^2}\, \right)\,.
\label{R}
\end{eqnarray}
Here $b=\frac{\sqrt{3}}{\sqrt{2}\,r_{0} } $ where $r_{0}$ is the glueball's {\it rms} radius  
and  ${\cal J}_{0}(x)=\sin x /x $. In (\ref{R}) one finds the following notation $\omega^{(i)}_{1}$
and $\omega^{(i)}_{2}$, where the index $i$ corresponds to the number of the evaluated diagram
in Fig. \ref{hfi_fig}. The glueball-glueball potential is shown in the Fig. \ref{graf_5}.

\vspace{-0.5cm}

The glueball-glueball potential is a short range and deep potential. However, it is interesting to
investigate if the gluenall glueball potential will form bound states.

\section*{Acknowledgments}

This work was partially supported by FAPERGS and CNPq.

\end{document}